\documentclass[aps,floatfix,nofootinbib,showpacs,twocolumn,superscriptaddress]{revtex4}

\usepackage{graphicx}
\usepackage{dcolumn}
\usepackage{amsmath}
\usepackage{longtable}
\usepackage{color}

\definecolor{purple}{rgb}{0.5,0,0.5}
\definecolor{blue}{rgb}{0.0,0,0.9}
\usepackage[colorlinks=true, pdfstartview=FitV, linkcolor=purple, citecolor= purple, urlcolor=blue]{hyperref}

\newcommand{\sfrac}[2]{\mbox{\footnotesize $\displaystyle \frac{#1}{#2}$}}

\begin{document}
\title{Vacuum pseudoscalar susceptibility}

\author{Lei Chang}
\affiliation{Institute of Applied Physics and Computational
Mathematics, Beijing 100094, China}

\author{Yu-Xin Liu}
%\email[Corresponding author: ]{ yxliu@pku.edu.cn}
\affiliation{Department of Physics, Peking University, Beijing 100871,
China}
\affiliation{State Key Laboratory of Nuclear Physics and Technology, Peking University, Beijing 100871, China}
\affiliation{Center of Theoretical Nuclear Physics, National
Laboratory of Heavy Ion Accelerator, Lanzhou 730000, China}

\author{Craig D.\ Roberts}
\affiliation{Department of Physics, Peking University, Beijing 100871, China}
\affiliation{Physics Division, Argonne National Laboratory, Argonne, Illinois 60439, USA}

\author{Yuan-Mei Shi}
\affiliation{Department of Physics, Nanjing Xiaozhuang College,
Nanjing 211171, China}

\author{Wei-Min Sun}
\affiliation{Department of Physics, Nanjing University, Nanjing
210093, China} \affiliation{Joint Center for Particle, Nuclear
Physics and Cosmology, Nanjing 210093, China}

\author{Hong-Shi Zong}
\affiliation{Department of Physics, Nanjing University, Nanjing
210093, China} \affiliation{Joint Center for Particle, Nuclear
Physics and Cosmology, Nanjing 210093, China}

\date{\today}

\begin{abstract}
We derive a novel model-independent result for the pion susceptibility in QCD via the isovector-pseudoscalar vacuum polarisation.  In the neighbourhood of the chiral-limit, the pion susceptibility can be expressed as a sum of two independent terms.  The first expresses the pion-pole contribution.  The second is identical to the vacuum chiral susceptibility, which describes the response of QCD's ground-state to a fluctuation in the current-quark mass.  In this result one finds a straightforward explanation of a mismatch between extant estimates of the pion susceptibility.
\end{abstract}


\maketitle

Colour-singlet current-current correlators or, equivalently, the associated vacuum polarisations, play an important role in QCD because they are directly related to observables.  The vector vacuum polarisation, e.g., couples to real and virtual photons.  It is thus basic to the analysis and understanding of the process $e^+ e^- \to\,$hadrons \cite{Gonsalves:2008rd,Krein:1990sf}.  In addition, analysis of the large Euclidean-time behaviour of a carefully chosen correlator can yield a hadron's mass \cite{Morningstar:2005pv,Bhagwat:2007rj}; and correlators are also amenable to analysis via the operator product expansion and are therefore fundamental in the application of QCD sum rules \cite{Colangelo:2000dp}.

In the latter connection, the vacuum pseudoscalar susceptibility (also called the pion susceptibility) plays a role in the sum-rules estimate of numerous meson-hadron couplings; e.g., the strong and parity-violating pion-nucleon couplings, $g_{\pi N N}$ and $f_{\pi N N}$, respectively \cite{Reinders:1982hd,Henley:1995ad,Johnson:1997gq}.  Furthermore, as will become plain herein, the pion susceptibility is as intimate a probe of QCD's vacuum structure as the scalar susceptibility \cite{Chang:2008ec} but its veracious analysis is more subtle, with conflicts and misconceptions being common \cite{Henley:1995ad,Johnson:1997gq,Belyaev:1983hn,Chanfray:2001tf,Zong:2003kf}.

We approach the vacuum pseudoscalar susceptibility via the isovector-pseudoscalar vacuum polarization, which can be written\footnote{In our Euclidean metric:  $\{\gamma_\mu,\gamma_\nu\} = 2\delta_{\mu\nu}$; $\gamma_\mu^\dagger = \gamma_\mu$; $\gamma_5= \gamma_4\gamma_1\gamma_2\gamma_3$; $a \cdot b = \sum_{i=1}^4 a_i b_i$; and $P_\mu$ timelike $\Rightarrow$ $P^2<0$.}
\begin{equation}
\label{omega5}
\omega_5^{ij}(P;\zeta) =   N_c\, {\rm tr} Z_4 \!\int_q^\Lambda\!
\sfrac{i}{2} \gamma_{5} \tau^i S(q_+) i\Gamma_{5}^j(q;P) S(q_-)\,,
\end{equation}
where the trace is over flavour and spinor indices; $\zeta$ is the renormalisation scale; $Z_4(\zeta,\Lambda)$ is the Lagrangian mass-term renormalisation constant, which depends implicitly on the gauge parameter;\footnote{Physical quantities obtained from Eq.\,(\ref{omega5}) are manifestly gauge-invariant.} and $\int_q^\Lambda:= \int^\Lambda d^4 q/(2\pi)^4$ represents a symmetry-preserving regularisation of the integral, with $\Lambda$ the regularisation mass-scale which is taken to infinity as the last step in a complete calculation.

Herein we will subsequently assume isospin symmetry; viz., equal $u$- and $d$-quark current-masses, in considering the isovector-channel.  An extension to three flavours and the flavour-singlet channel can be pursued following the methods of Ref.\,\cite{Bhagwat:2007ha}.

In Eq.\,(\ref{omega5}), $S$ is the dressed-quark propagator and $\Gamma_{5}$ is the fully-dressed pseudoscalar vertex, both of which depend on the renormalisation point.  The propagator is obtained from QCD's gap equation; namely,
\begin{equation}
\begin{array}{rcl}
S(p)^{-1} &=& \displaystyle Z_2 \,(i\gamma\cdot p + m^{\rm bm}) + \Sigma(p)\,, \label{gendse} \\
\Sigma(p)&=& \displaystyle Z_1 \int^\Lambda_q\! g^2 D_{\mu\nu}(p-q) \frac{\lambda^a}{2}\gamma_\mu S(q) \frac{\lambda^a}{2}\Gamma_\nu(q,p) , %\label{gensigma}
\end{array}
\end{equation}
where $D_{\mu\nu}(k)$ is the dressed-gluon propagator, $\Gamma_\nu(q,p)$ is the dressed-quark-gluon vertex, and $m^{\rm bm}$ is the $\Lambda$-dependent $u$- and $d$-quark current-quark bare mass.  The quark-gluon-vertex and quark wave-function renormalisation constants, $Z_{1,2}(\zeta,\Lambda)$, also depend on the gauge parameter.  %NB.\ We envisage use of a mass-independent renormalisation scheme.

The gap equation's solution has the form
\begin{eqnarray}
%\nonumber
 S(p)^{-1} & = & i \gamma\cdot p \, A(p^2;\zeta^2) + B(p^2;\zeta^2) %\\
%
%& =& \frac{1}{Z(p^2,\zeta^2)}\left[ i\gamma\cdot p + M(p^2,\zeta^2)\right] .
\label{sinvp}
\end{eqnarray}
and the mass function $M(p^2)=B(p^2,\zeta^2)/A(p^2,\zeta^2)$ is renormalisation point independent.  The propagator is obtained from Eq.\,(\ref{gendse}) augmented by a renormalisation condition.

Since QCD is asymptotically free, the chiral limit is defined by
\begin{equation}
Z_2(\zeta,\Lambda) \, m^{\rm bm}(\Lambda) \equiv 0\,,\; \forall \Lambda \gg \zeta\,,
\end{equation}
which is equivalent to requiring that the renormalisation point invariant current-quark mass is zero; i.e., $\hat m = 0$.  A mass-independent renormalisation scheme can then be implemented by fixing all renormalisation constants in the chiral limit \cite{Weinberg:1951ss}; namely, one solves the chiral limit gap equation subject to the requirement
\begin{equation}
\label{renormS} \left.S^{-1}_{\hat m=0}(p)\right|_{p^2=\zeta^2} = i\gamma\cdot p \,.
%+ m_f(\zeta)\,,
\end{equation}
This is implicit in the subsequent analysis.  We note that
\begin{equation}
Z_2(\zeta,\Lambda) \, m^{\rm bm}(\Lambda)=Z_4(\zeta,\Lambda) \, m(\zeta)\,,
\end{equation}
where $m(\zeta)$ is the familiar running current-quark mass.

The pseudoscalar vertex is determined from an inhomogeneous Bethe-Salpeter equation; viz.,
\begin{equation}
[\Gamma_5^j(k;P)]_{tu} = Z_4 [\sfrac{1}{2}\gamma_5\tau^j]_{tu} + \int_q^\Lambda\! [\chi_5^j(q;P)]_{sr} K_{tu}^{rs}(q,k;P)\,,
\end{equation}
where $k$ is the relative- and $P$ the total-momentum of the quark-antiquark pair; $r$, $s$, $t$, $u$ represent colour, flavour and spinor indices;
\begin{equation}
\chi_5^j(k;P)=S(k_+) \Gamma_5^j(k;P) S(k_-)\,,
\end{equation}
$k_\pm = k\pm P/2$, without loss of generality owing to the symmetry-preserving nature of the regularisation scheme; and $K(q,k;P)$ is the fully-amputated two-particle-irreducible quark-antiquark scattering kernel.

Much of the preceding material recapitulates results familiar from QCD's Dyson-Schwinger equations (DSEs) \cite{Roberts:1994dr,Roberts:2007jh}, which also provide the foundation for our subsequent analysis.  Consider, then, that in the presence of a spacetime-independent pseudoscalar source, $\vec{s}_5\neq 0$, associated with a term
\begin{equation}
\int d^4 x\, \bar q(x)  \, \sfrac{i}{2} \gamma_5 \vec{\tau}\cdot \vec{s}_5 \, q(x)
\end{equation}
in the action, one can define a vacuum pseudoscalar condensate, whose gauge-invariant, properly renormalised form in QCD is
\begin{equation}
\label{sigma5}
\sigma_5^j(\vec{s}_5,m;\zeta,\Lambda) = Z_4 N_c {\rm tr} \! \int_q^\Lambda\! \sfrac{i}{2}\gamma_5 \tau^j S(q;\vec{s}_5,m;\zeta)\,.
\end{equation}
This is analogous to the vacuum quark condensate \cite{Langfeld:2003ye}
\begin{equation}
\label{sigma}
\sigma(m;\zeta,\Lambda) = Z_4 N_c {\rm tr}\! \int_q^\Lambda\! \sfrac{1}{2} \tau^0 \, S(q;m;\zeta)\,,
\end{equation}
$\tau^0 = {\rm diag}[1,1]$, whose source-term in the QCD action is that associated with the current-quark mass.

It should be emphasised that when $\hat m=0$, it is only foreknowledge of nonzero current-quark masses via the Higgs mechanism which leads one to express dynamical chiral symmetry breaking (DCSB) as
\begin{equation}
\label{sigmam}
-\langle \bar q q \rangle_\zeta^0 = \lim_{m\to 0}\sigma(m;\zeta,\Lambda) \neq 0\,,
\end{equation}
instead of choosing a different vacuum vector; e.g., $(\sigma;\vec{s}_5) \propto (0;1,-i,0)$.
%Following was modified ... anomaly knows about a U(1) chiral rotation, which means it knows about parity.
Moreover, Eq.\,(\ref{sigmam}) defines what we mean by an isoscalar-scalar configuration: isovector-pseudoscalar correlations are by convention measured with respect to this configuration.

These observations highlight the importance of the pseudoscalar susceptibility
\begin{equation}
{\cal X}_5^{ij}(\zeta):= \left. \frac{\partial}{\partial s_5^i}\sigma_5^j(\vec{s}_5,m;\zeta,\Lambda) \right|_{{\vec s}_5=0}.
\end{equation}
Following the method of Ref.\,\cite{Chang:2008ec}, it is straightforward to show that
\begin{equation}
\label{X5}
{\cal X}_5^{ij}(\zeta) = - 2\, \omega_5^{ij}(P=0;\vec{s}_5=0,\hat m;\zeta)\,.
\end{equation}
NB.\ Whilst hitherto we have not specified a regularisation procedure for the susceptibility, it can rigorously be defined via a Pauli-Villars procedure \cite{Chang:2008ec}.

We are interested in the value of ${\cal X}_5^{ij}(\zeta)$ in the neighbourhood of the chiral limit.  Therein one may write \cite{Maris:1997hd}
\begin{equation}
\label{G5k0}
i \Gamma_5^j(k;0) = \sfrac{1}{2} i\gamma_5 \tau^j E_5^R(k;0) + \frac{r_\pi}{m_\pi^2} \Gamma_\pi^j(k;0)\,,
\end{equation}
where $\Gamma_\pi^j(k;P)$ is the pion bound-state's canonically-normalised Bethe-Salpeter amplitude; $r_\pi(\zeta)$, determined by
\begin{eqnarray}
i \delta^{jk} r_\pi(\zeta) & = & \langle 0 | \bar q \,\sfrac{1}{2}\gamma_5 \tau^k \, q | \pi^j \rangle \\
& = &N_c {\rm tr} Z_4 \int_q^\Lambda \sfrac{1}{2} \gamma_5 \tau^k \chi_\pi^j(q;P)\,,
\end{eqnarray}
is the residue of this bound-state in the inhomogeneous pseudoscalar vertex; and $E_5^R(k;P)$ is a part of the inhomogeneous pseudoscalar vertex which is regular as \mbox{$P^2 +m_\pi^2 \to 0$}.

These statements will not be surprising once one recalls that the solution of a linear, inhomogeneous integral equation is a sum; viz., the regular solution of the inhomogeneous equation plus a solution of the homogeneous equation, here, naturally, the canonically normalised solution.  In terms of this solution, the pion's leptonic decay constant is expressed through
\begin{eqnarray}
\delta^{jk} f_\pi\, P_\mu & = & \langle 0 | \bar q\, \sfrac{1}{2}\gamma_5\gamma_\mu \tau^k \, q | \pi^j \rangle \\
& = &N_c {\rm tr} Z_2 \int_q^\Lambda \sfrac{1}{2} \gamma_5 \gamma_\mu \tau^k \chi_\pi^j(q;P)\,.
\end{eqnarray}

One can turn to the axial-vector Ward-Takahashi identity in order to determine $E_5^R(k;P)$; viz.,
\begin{eqnarray}
\nonumber
&& P_\mu \Gamma_{5\mu}^j(k;P) + \,2 m(\zeta) \,i\Gamma_5^{j}(k;P)\\
&=& S^{-1}(k_+) \sfrac{1}{2} i\gamma_5 \tau^j +  \sfrac{1}{2} i\gamma_5 \tau^j S^{-1}(k_-) \,,
\label{avwtim}
\end{eqnarray}
where $\Gamma_{5\mu}^j(k;P)$ is the inhomogeneous axial-vector vertex.  At $P=0$ with $\hat m\neq 0$ there is no pole contribution on the left-hand-side and hence Eq.\,(\ref{avwtim}) states
\begin{equation}
\label{mE5Bm}
 m(\zeta) \,E_5^R(k;P=0) = B(k^2;m;\zeta^2)\,,
\end{equation}
namely, this regular piece of the pseudoscalar vertex is completely determined by the scalar part of the $\hat m\neq 0$ quark self-energy.  NB.\ It is straightforward to verify Eq.\,(\ref{mE5Bm}), order-by-order, via the gap and Bethe-Salpeter equations using the systematic, nonperturbative, symmetry-preserving DSE truncation scheme introduced in Refs.\,\cite{Munczek:1994zz,Bender:1996bb}.

We insert Eq.\,(\ref{G5k0}) into Eq.\,(\ref{X5}) to obtain
\begin{eqnarray}
{\cal X}_5^{ij}(\zeta) & \stackrel{\hat m ~ 0}{=} & \delta^{ij} {\cal X}_5(\zeta)\,,\\
{\cal X}_5(\zeta) & = & {\cal X}_5^\pi(\zeta) + {\cal X}_5^R(\zeta)+ \mbox{O}(\hat m) \,;
\label{X52}
\end{eqnarray}
namely, in the neighbourhood of $\hat m =0$ the susceptibility splits into a sum of two terms.  The first of these expresses the contribution of the pion pole, $\mbox{O}(\hat m^{-1})$, and can readily be expressed in a closed form
\begin{equation}
{\cal X}_5^\pi(\zeta)= \frac{2 \, r_\pi(\zeta)^2}{m_\pi^2}
 \stackrel{\hat m=0}{=}  - \, \frac{\langle \bar q q\rangle^0_\zeta}{m(\zeta)} \,,
\end{equation}
where the last equality is proved in Ref.\,\cite{Maris:1997hd}.
The second term in Eq.\,(\ref{X52}), $ \mbox{O}(\hat m^{0})$, is implicitly determined via
\begin{eqnarray}
\nonumber
\lefteqn{ m(\zeta) \, {\cal X}_5^R(\zeta) \, \delta^{jk}}\\
\nonumber
&\stackrel{\hat m \sim 0}{=}&  - N_c\, {\rm tr} Z_4 \!\int_q^\Lambda\!
i \gamma_{5} \tau^k S(q) \sfrac{i}{2} \gamma_{5} \tau^j B(q^2;m) S(q)\\
&=& \delta^{jk}\, \sigma(m;\zeta,\Lambda)\,, \label{X5R}
\end{eqnarray}
where the last line is obtained using $\{\gamma_5,\gamma_\mu\} =0$.% and the massive scalar vacuum quark condensate is expressed in Eq.\,(\ref{sigma}).

We can now proceed to our desired result.  Equation\,(\ref{X5R}) entails
\begin{equation}
{\cal X}_5^R(\zeta;m) = {\cal X}(\zeta) + \mbox{O}(\hat m)\,,
\end{equation}
where the vacuum chiral susceptibility is \cite{Chang:2008ec}
\begin{equation}
{\cal X}(\zeta) = \left. \frac{\partial}{\partial m(\zeta)}\, \sigma(m;\zeta,\Lambda) \right|_{\hat m=0}.
\end{equation}
Hence we arrive at a model-independent consequence of chiral symmetry and the pattern by which its broken in QCD; namely,
\begin{equation}
\label{main}
{\cal X}_5(\zeta) \stackrel{\hat m ~ 0}{=}  - \, \frac{\langle \bar q q\rangle^0_\zeta}{m(\zeta)}  + {\cal X}(\zeta) + \mbox{O}(\hat m)\,.
\end{equation}

\begin{table}[t]
\caption{\label{tableX5}
Vacuum pseudoscalar susceptibility and related quantities, computed using the two kernels of the Bethe-Salpeter equation described in connection with Eqs.\,(\protect\ref{IRGs}), (\protect\ref{rainbowV} and (\protect\ref{bcvtx}).
Dimensioned quantities are listed in GeV, $\kappa := -(\langle \bar q q \rangle^0_\zeta)^{1/3}$ and $f_\pi^0$ is the pion's chiral-limit leptonic decay constant.
The entries were compiled from Refs.\,\protect\cite{Chang:2008ec,Chang:2009zb}.
NB.\ For quantitative comparison with some other studies \protect\cite{Henley:1995ad,Johnson:1997gq,Belyaev:1983hn,Zong:2003kf}, our results for ${\cal X}$ should be multiplied by $(2\pi)^2$.
}
\begin{center}
\begin{tabular*}%{|c|c|c|c|c|c|c|}\hline
{\hsize}
{
c@{\extracolsep{0ptplus1fil}}
c@{\extracolsep{0ptplus1fil}}
c@{\extracolsep{0ptplus1fil}}
|c@{\extracolsep{0ptplus1fil}}
c@{\extracolsep{0ptplus1fil}}
c@{\extracolsep{0ptplus1fil}}
|c@{\extracolsep{0ptplus1fil}}
c@{\extracolsep{0ptplus1fil}}}
Vertex & $\sqrt{D}$ & $\omega\;\;$ & $\kappa$  & $f_\pi^0$ & $m$ \rule{0em}{0ex} & $\sqrt{{\cal X}_5^\pi}$ & $\sqrt{{\cal X}_5^R}$\\\hline
% 3.125 GeV^2  0.152 GeV^2
\rule{0em}{3ex} RL [Eq.\,(\protect\ref{rainbowV})] & 1& $\sfrac{1}{2}\;\;$ & 0.25 & 0.091 & 0.0050 \rule{0em}{0ex} & 1.77 & 0.39 \\
% 2.746 GeV^2  0.0784 GeV^2
\rule{0em}{4ex} BC [Eq.\,(\protect\ref{bcvtx})] & \sfrac{1}{\surd 2} & $\sfrac{1}{2}\;\;$ & 0.26 & 0.11\;\; & 0.0064 \rule{0em}{0ex} & 1.66 & 0.28 \\\hline
\end{tabular*}
\end{center}
\end{table}

For illustration, in Table\,\ref{tableX5} we report numerical values computed from two models for the gap equation's kernel.  Namely, we simplify the renormalisation-group-improved effective interaction in Ref.\,\cite{Maris:1997tm}
\begin{eqnarray}
\nonumber \lefteqn{Z_1 g^2 D_{\rho \sigma}(p-q) \Gamma_\sigma^a(q,p)} \\
& =& {\cal G}((p-q)^2) \, D_{\rho\sigma}^{\rm free}(p-q) \frac{\lambda^a}{2}\Gamma_\sigma(q,p)\,, \label{KernelAnsatz}
\end{eqnarray}
wherein $D_{\rho \sigma}^{\rm free}(p-q)$ is the Landau-gauge free gauge-boson propagator, through the choice
\begin{equation}
\label{IRGs}
\frac{{\cal G}(s)}{s} = \frac{4\pi^2}{\omega^6} \, D\, s\, {\rm e}^{-s/\omega^2},
\end{equation}
which is a finite width representation of the form introduced in Ref.\,\cite{Munczek:1983dx}.  This interaction has been rendered as an integrable regularisation of $1/k^4$ \cite{McKay:1996th}.  Equation~(\ref{IRGs}) delivers an ultraviolet finite model gap equation.  Hence, the regularisation mass-scale can be removed to infinity and the renormalisation constants set equal to one.  %That is also true of Eq.\,(\ref{chipractical}).

The kernel is completed by specifying the dressed-quark-gluon vertex.  At leading-order in the systematic DSE truncation scheme \cite{Munczek:1994zz,Bender:1996bb} the vertex is
\begin{equation}
\label{rainbowV}
\Gamma_\sigma(q,p) = \gamma_\sigma\,.
\end{equation}
This defines the rainbow-ladder (RL) truncation.  One can alternatively employ \emph{Ans\"atze} for the vertex whose diagrammatic content is unknown.  A class of such models, which has seen much use in diverse applications; e.g., Refs.\,\cite{Chang:2008sp,Chang:2008ec,Ivanov:2007cw,Chen:2008zr,Cloet:2008re,%
Chang:2009zb}, can be characterised by \cite{Ball:1980ay}
\begin{eqnarray}
\label{bcvtx}
\nonumber \lefteqn{i\Gamma_\sigma(k,\ell)  =
i\Sigma_A(k^2,\ell^2)\,\gamma_\sigma + (k+\ell)_\sigma }\\
&\times &
\left[\frac{i}{2}\gamma\cdot (k+\ell) \,
\Delta_A(k^2,\ell^2) + \Delta_B(k^2,\ell^2)\right] \!,
\end{eqnarray}
where
\begin{eqnarray}
\Sigma_F(k^2,\ell^2)& = &\frac{1}{2}\,[F(k^2)+F(\ell^2)]\,,\;\\
\Delta_F(k^2,\ell^2) &=&
\frac{F(k^2)-F(\ell^2)}{k^2-\ell^2}\,,
\label{DeltaF}
\end{eqnarray}
with $F= A, B$; viz., the scalar functions in Eq.\,(\ref{sinvp}).  This \emph{Ansatz} satisfies the vector Ward-Takahashi identity and is often referred to as the BC vertex.

Equation\,(\ref{main}) is a remarkable result, which is nonetheless readily understood.  Recall that in the absence of a current-quark mass, the two-flavour action has a SU$_L(2)\otimes\,$SU$_R(2)$ symmetry; and, moreover, that ascribing scalar-isoscalar quantum numbers to the QCD vacuum is a convention contingent upon the form of the current-quark mass term.

It follows that the massless action cannot distinguish between the continuum of sources specified by
\begin{equation}
{\rm constant} \times \int d^4 x\,  \bar q(x) \, {\rm e}^{i \gamma_5 \vec{\tau}\cdot \vec{\theta}} q(x)\,,\; |\theta|\in[0,2\pi)\,.
\end{equation}
Hence, the regular part of the vacuum susceptibility must be identical when measured as the response to any one of these sources, so that ${\cal X}_R={\cal X}$ for all choices of $\vec{\theta}$.  This is the content of the so-called ``Mexican hat'' potential, which is used in building models for QCD.
The magnitude of ${\cal X}$ depends on whether the chiral symmetry is dynamically broken, or not; and the strength of the interaction as measured with respect to the critical value required for DCSB \cite{Chang:2008ec}.  When the symmetry is dynamically broken, then the Goldstone modes appear, by convention, in the pseudoscalar-isovector channel, and thus the pole contributions appear in ${\cal X}_5$ but not in the chiral susceptibility.  It is valid to draw an analogy with the Weinberg sum rule \cite{Weinberg:1967kj,Chang:2008sp}.

With Eq.\,(\ref{main}) we have, in addition, provided a novel, model-independent perspective on a mismatch between the evaluation of the pion susceptibility using either a two-point or three-point sum rule.  Namely, the two-point study \cite{Belyaev:1983hn} produces the pion pole contribution, ${\cal X}_5^\pi$, which is also the piece emphasised in Ref.\,\cite{Chanfray:2001tf}, whereas a three-point method \cite{Johnson:1997gq} isolates the regular piece, ${\cal X}_5^R$, because a vacuum saturation \emph{Ansatz} is implemented in the derivation.  Thus, the analyses are not essentially in conflict.  Instead, they emphasise different, independent pieces of the susceptibility, which, with care, can be distinguished.  However, in a sum-rules estimate of pion-nucleon coupling constants, only the regular piece should be retained \cite{Reinders:1982hd}.
%
% Compare Eq.(12) of Reinders et al. with our Eq.\,(\ref{X5R}).

We note in closing that the vacuum tensor susceptibilities, which can be related to the nucleon's tensor charges \cite{He:1996wy}, can similarly be analysed.  Such a study is underway.

%\bigskip

\begin{acknowledgments}
This work was supported by:
the National Natural Science Foundation of China, under Contract Nos.~10425521, 10675007, 10705002, 10775069 and 10935001;
the Major State Basic Research Development Program under contract No.~G2007CB815000;
and the United States Department of Energy, Office of Nuclear Physics, contract no.~DE-AC02-06CH11357.
\end{acknowledgments}

\bibliography{LCPSus}

\end{document}